\documentclass[conference, 10pt]{IEEEtran}
\IEEEoverridecommandlockouts
\usepackage[utf8x]{inputenc}
\usepackage[T1]{fontenc}
\usepackage[hidelinks]{hyperref}
\usepackage{babel}
\usepackage{cite}
\usepackage{amsmath,amssymb,amsfonts,mathtools,gensymb}
\usepackage{algorithmic}
\usepackage{graphicx}
\usepackage{textcomp}
\usepackage[x11names]{xcolor}
\def\BibTeX{{\rm B\kern-.05em{\sc i\kern-.025em b}\kern-.08em
		T\kern-.1667em\lower.7ex\hbox{E}\kern-.125emX}}
\usepackage[final]{microtype}

\definecolor[named]{LightKellyGreen}{HTML}{B2F07F}
\definecolor[named]{LightUPCblue}{HTML}{5CBEFF}

\usepackage[toc, acronym]{glossaries}
\glsdisablehyper

\usepackage{siunitx}
\usepackage{makecell}
\usepackage{graphicx}
\graphicspath{{figures/}}
\usepackage{enumitem}
\usepackage{siunitx}
\usepackage{booktabs}

\usepackage{tikz}
\usepackage{circuitikz}
\usepackage{pgfplots}
\pgfplotsset{compat=1.16}
\pgfplotsset{compat=newest}
\usepgfplotslibrary{groupplots}
\usepgfplotslibrary{polar}
\usepgfplotslibrary{smithchart}
\usepgfplotslibrary{statistics}
\usepgfplotslibrary{dateplot}
\usepgfplotslibrary{ternary}
\usetikzlibrary{arrows.meta}
\usetikzlibrary{backgrounds}
\usetikzlibrary{shapes.misc}
\usetikzlibrary{angles, quotes}
\usetikzlibrary{decorations.pathmorphing}
\usetikzlibrary{calc}
\usetikzlibrary{shapes.geometric}
\tikzset{>=latex}
\usepgfplotslibrary{patchplots}
\usepgfplotslibrary{fillbetween}

\usepackage[nameinlink,capitalize]{cleveref}
\crefname{figure}{Fig.}{Figs.}
\crefname{equation}{}{}
\crefname{section}{Sec.}{Secs.}

\newcommand{\ie}{\textit{i.e. }}

\newcommand{\cov}[2][]{\ensuremath{\mathbf{C}_{#2}^{\mathrm{#1}}}}
\newcommand{\norm}[2]{\ensuremath{\left\|#1\right\|_{\mathrm{#2}}}}
\newcommand{\herm}[1]{\ensuremath{#1^{\mathrm{H}}}}
\newcommand{\trans}[1]{\ensuremath{#1^{\mathrm{T}}}}
\newcommand{\expec}[2][]{\ensuremath{\mathrm{E}_{#1}\left[#2\right]}}
\newcommand{\trace}[1]{\ensuremath{\mathrm{tr}\left(#1\right)}}
\newcommand{\euler}{\ensuremath{\mathrm{e}}}
\newcommand{\im}{\ensuremath{\mathrm{j}}}

\newcommand{\id}[1]{\ensuremath{\mathbf{I}_{#1}}}
\newcommand{\bsf}[2][]{\ensuremath{\boldsymbol{\mathsf{#2}}^{\mathrm{#1}}}}
\newcommand{\cnumbers}{\ensuremath{\mathbb{C}}}

\newcommand{\N}[1]{\ensuremath{N_{\mathrm{#1}}}}

\newcommand{\complexnormal}{\ensuremath{\mathcal{C\!N}}}
\newcommand{\zeros}[1]{\ensuremath{\mathbf{0}_{#1}}}

\newcommand{\dist}[2][]{\ensuremath{d_{#2}^{\mathrm{#1}}}}
\newcommand{\aresp}[1]{\ensuremath{\mathbf{a}}_{\mathrm{#1}}}

\newcommand{\harmn}[2]{\ensuremath{H_{#1}^{(#2)}}}

\newcommand{\R}[1]{\ensuremath{R_{\mathrm{#1}}}}

\newcommand{\Z}[1]{\ensuremath{Z_{\mathrm{#1}}}}
\newcommand{\bfZ}[2][]{\ensuremath{\mathbf{Z}_{\text{#2}}^{\text{#1}}}}
\newcommand{\bsfZ}[2][]{\ensuremath{\boldsymbol{\mathsf{Z}}_{\text{#2}}^{\text{#1}}}}

\newcommand{\bsfz}[2][]{\ensuremath{\boldsymbol{\mathsf{z}}_{\text{#2}}^{\text{#1}}}}
\newcommand{\bsfvs}[1]{\ensuremath{\boldsymbol{\mathsf{v}}_{\mathrm{#1}}}}
\newcommand{\sfvs}[1]{\ensuremath{\mathsf{v}}_{\mathrm{#1}}}
\newcommand{\bsfis}[1]{\ensuremath{\boldsymbol{\mathsf{i}}_{\mathrm{#1}}}}
\newcommand{\sfis}[1]{\ensuremath{\mathsf{i}}_{\mathrm{#1}}}

\let\originalleft\left
\let\originalright\right
\renewcommand{\left}{\mathopen{}\mathclose\bgroup\originalleft}
\renewcommand{\right}{\aftergroup\egroup\originalright}

\begin{document}
	\newacronym{bs}{BS}{base station}
\newacronym{ehf}{EHF}{extremely high frequency}
\newacronym{lna}{LNA}{low-noise amplifier}
\newacronym{los}{LoS}{line-of-sight}
\newacronym{mimo}{MIMO}{multiple-input multiple-output}
\newacronym{miso}{MISO}{multiple-input single-output}
\newacronym{nlos}{NLoS}{non-line-of-sight}
\newacronym{simo}{SIMO}{single-input multiple-output}
\newacronym{snr}{SNR}{signal-to-noise ratio}
\newacronym{thf}{THF}{tremendously high frequency}
\newacronym{ue}{UE}{user equipment}
\newacronym{ula}{ULA}{uniform linear array}

	\bibliographystyle{IEEEtran-normspace}
	\bstctlcite{IEEEexample:BSTcontrol}

	\title{On the Unilateral Approximation Condition with Linear Arrays in Near-Field NLoS Propagation\\
		\thanks{This work was supported by project MAYTE (PID2022-136512OB-C21) by MICIU/AEI/10.13039/501100011033 and ERDF/EU, grants 2021 SGR 01033 and 2021 SGR 00603 by Departament de Recerca i Universitats de la Generalitat de Catalunya and grant 2023 FI ``Joan Oró'' 00050 by Departament de Recerca i Universitats de la Generalitat de Catalunya and the ESF+.
        L.~Sanguinetti was supported in part by the Italian Ministry of Education and Research (MUR) in the framework of the FoReLab Project (Department of Excellence).}
	}
	\author{
		\IEEEauthorblockN{
			Aniol Martí\IEEEauthorrefmark{1}, Luca Sanguinetti\IEEEauthorrefmark{2}, and Xavier~Gràcia\IEEEauthorrefmark{3}
		}
		\IEEEauthorblockA{
			\textit{\IEEEauthorrefmark{1}Department of Signal Theory and Communications, Universitat Politècnica de Catalunya}\\
		}
		\IEEEauthorblockA{
			\textit{\IEEEauthorrefmark{2}Dipartimento di Ingegneria dell'Informazione, University of Pisa}\\
		}
        \IEEEauthorblockA{
			\textit{\IEEEauthorrefmark{3}Department of Mathematics, Universitat Politècnica de Catalunya}\\
		}
        \IEEEauthorblockA{
            aniol.marti@upc.edu, luca.sanguinetti@unipi.it, xavier.gracia@upc.edu
        }
	}

	\maketitle

	\begin{abstract}
        This paper investigates the validity of the unilateral approximation in near-field \acrfull{nlos} propagation for large-scale linear antenna arrays.
        Utilizing multiport communication theory, the study evaluates whether the electromagnetic interaction from the receiver back to the transmitter can be neglected.
        The analysis examines both discrete arrays with fixed spacing and continuous arrays with fixed aperture as the number of antennas increases.
        Results show the approximation is asymptotically satisfied on average for both cases.
        This provides a theoretical foundation for using correlated Rayleigh fading models in physically consistent next-generation wireless systems.
	\end{abstract}


	\section{Introduction}
        \IEEEPARstart{T}{he increasing} demand for high data rates in modern wireless communication systems, together with the enormous growth in the number of connected devices, has driven new wireless communication standards toward extremely large antenna arrays and operation in the \acrfull{ehf} and \acrfull{thf} bands~\cite{chafii_twelve_2023}.

        Both the use of extremely large arrays and the shift to mmWave and even higher frequency bands challenge the propagation models traditionally employed in classical \acrfull{mimo} systems, which rely on the assumptions of uncoupled antennas operating in the far field and channels modeled as Rayleigh or Rician fading.
        On the one hand, operation at such high frequencies increases the far-field distance to several kilometers, so newly proposed communication schemes should be analyzed under near-field propagation conditions~\cite{marti_harnessing_2025}.
        On the other hand, the use of massive numbers of densely packed antennas, such as those in holographic \acrshort{mimo}, inevitably gives rise to mutual coupling between antenna elements~\cite[Sec.~8.7]{balanis_antenna_2016}.

        Evaluating the performance of these next-generation wireless communication systems therefore requires physically consistent propagation models capable of accounting for the electromagnetic interaction between antennas, either within the transmitter or receiver arrays (intra-array coupling) or between antennas at both ends of the communication link (inter-array coupling)~\cite{ivrlac_toward_2010}.
        Hence, it becomes necessary to develop mathematically tractable models that remain physically accurate while preserving the abstraction level required by the signal processing and communications community.

        Fortunately, several approaches to address these challenges have been proposed~\cite{mezghani_reincorporating_2024,zhu_electromagnetic_2024}.
        Among them, \textit{multiport communication theory} has emerged as a particularly promising framework~\cite{ivrlac_toward_2010}.
        This approach adopts a circuit-theoretic perspective in which the multiple-antenna communication system is modeled as a multiport black box characterized by its impedance or scattering matrix.

        Multiport theory has already been widely employed in the \acrshort{mimo} literature to analyze several aspects, including array gain~\cite{laas_limits_2020}, uplink/downlink reciprocity~\cite{damico_holographic_2024}, the impact of mutual coupling on holographic \acrshort{mimo} channel estimation~\cite{tang_channel_2026}, symbol detection~\cite{marti_coherent_2025}, and spectral efficiency~\cite{lu_compact_2026}.
        Nevertheless, most of these works are focused on far-field communication so unilateral coupling networks are considered~\cite[Sec.~11.2]{pozar_microwave_2012}.
        A notable exception is~\cite{laas_limits_2020}, where the authors studied the array gain of a massive \acrshort{mimo} system operating in the near field and argued that unilateral networks are no longer physically consistent in such scenarios.
        However, no formal proof for the validity of the unilateral approximation in the near field was provided.

        In~\cite{marti_asymptotic_2024}, the authors employed the multiport framework to study the conditions under which the coupling induced by the receiver on the transmitter antennas can be neglected, \ie the unilateral approximation, under \acrfull{los} propagation conditions.

        In this paper, we extend these results to \acrfull{nlos} propagation, a highly relevant scenario in wireless communications.
        The analysis relies on a physically consistent model valid for coupled antennas and near-field propagation in order to identify the conditions under which the Rayleigh fading model can still be employed in the analysis of massive \acrfull{simo} and massive \acrfull{miso} systems.

    \section{System Model}
        \label{sec:system_model}
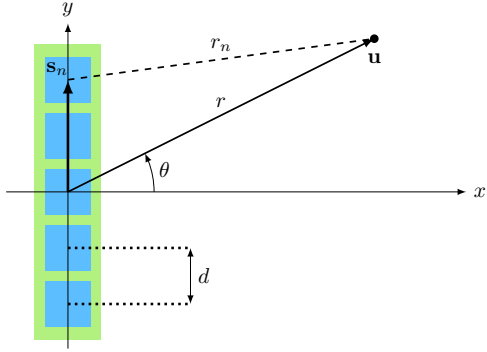
\begin{figure}[ht!]
            \centering
            \resizebox{0.75\columnwidth}{!}{%
                \begin{tikzpicture}
	% Variables
	\def\antsize{0.75};
	\def\dist{0.22*\antsize}
	\def\spacing{(\antsize+\dist};
	\def\N{2};
	\def\width{(\antsize+2*\dist}
	\def\height{\N*\spacing + 0.5*\antsize -0.5*\dist};
	\coordinate (ue) at (5, 2.5);
	\coordinate (orig) at (0, 0);

	% Array
	\fill[LightKellyGreen] (-0.5*\width, \height) rectangle (0.5*\width, -\height);
	% Elements
	\foreach \i in {\numexpr-\N\relax,...,\numexpr\N\relax}{
		\fill[LightUPCblue] (-0.5*\antsize, -0.5*\antsize+\i*\spacing) rectangle (0.5*\antsize, 0.5*\antsize+\i*\spacing);
	}

	% Labels
	\draw[dotted, very thick, black] (0, {(-\N+1)*\spacing}) -- (2, {(-\N+1)*\spacing});
	\draw[dotted, very thick, black] (0, {(-\N)*\spacing}) -- (2, {(-\N)*\spacing});
	\draw[<->, black] (2, {(-\N+1)*\spacing}) -- (2, {(-\N)*\spacing}) node[midway, right, black] {$d$};

	% Axis
	\draw[->, line cap=rect] (-1, 0) -- (6.5, 0) node (xaxis) [right] {$x$};
	\draw[->, line cap=rect] (0, -2.55) -- (0, 2.75) node[above] {$y$};

	% Vector u_n
	\draw[->, very thick] (orig) -- (0, 2*\spacing) node[left, black, xshift=4pt, yshift=5pt] {$\mathbf{s}_n$};

	% UE
	\fill[] (ue) circle (2pt);
	\draw[->, thick] (orig) -- (ue) node[midway, above] {$r$} node[right, below, yshift=-3pt] {$\mathbf{u}$};
	\pic [draw, ->, "$\theta$", angle eccentricity=1.15, angle radius=40] {angle = xaxis--orig--ue};
	\draw[-, thick, dashed] (0, 2*\spacing) -- (ue) node[midway, above, black] {$r_n$};
\end{tikzpicture}
            }
            \caption{Geometry of a \acrshort{ula} deployed along the $y$ axis.}
            \label{fig:array_model}
        \end{figure}

        We consider a communication system in \acrshort{nlos} propagation conditions, with a single-antenna \acrfull{ue} and a \acrfull{bs} with an $N$-element Hertzian dipole \acrfull{ula} deployed along the $y$ axis, as depicted in \cref{fig:array_model}.
        The size of the array is $D$, whereas the separation between antennas is $d$ and the wavelength is $\lambda$.
        Hence, the position of the $n$-th antenna with respect to the origin is $\mathbf{s}_n = \trans{(0, nd)}$, with $-(N-1)/2 \leq n \leq (N-1)/2$.
        The \acrshort{ue} is located at $\mathbf{u}=\trans{(x,y)}$.

        The discrete baseband representation of the signal received at the \acrshort{bs} is
        \begin{equation}
            \label{eq:signal_model}
            \bsf{y} = \bsf{h}\mathsf{x} + \bsf{z},\quad \bsf{y},\bsf{h},\bsf{z}\in\cnumbers^N,
        \end{equation}
        where $\bsf{z}\sim\complexnormal(\zeros{N},\cov{\bsf{z}})$ is additive Gaussian noise and $\mathsf{x}$ is the transmitted symbol.
        Traditionally, $\bsf{h}$ has been modeled as Rayleigh fading, or more generally as correlated Rayleigh fading to account for spatial correlation~\cite[Sec.~3.4]{heath_jr_foundations_2018}.
        However, this assumption relies on the scattering induced by the receiving array being negligible at the transmitter.
        Indeed, this has been shown to hold in the far field~\cite{ivrlac_toward_2010}, as well as in the near field under \acrshort{los} propagation~\cite{marti_asymptotic_2024}; yet, to the best of the authors' knowledge, no formal proof exists for the \acrshort{nlos} case.

        \subsection{Coupling Between Hertzian Dipoles}
            \label{sec:coupling_dipoles}
            The Hertzian dipole, also known as the infinitesimal dipole, is the most elementary antenna model satisfying Maxwell's equations.
            It consists of an electrically short and infinitesimally thin wire carrying an approximately uniform current density.
            Although Hertzian dipoles cannot be physically realized, they serve as fundamental building blocks for more complex antenna geometries~\cite[Sec.~4.2]{balanis_antenna_2016}.

            Mutual coupling is the electromagnetic interaction between nearby antenna elements, where the current from one source induces a voltage in another~\cite[Sec.~8.7]{balanis_antenna_2016}.
            Formally, this phenomenon can be characterized through the concept of \textit{mutual impedance}, defined as the ratio between the voltage induced in one antenna and the current flowing in another.

            Similarly, the \textit{self-impedance} of an antenna is the impedance measured at its terminals when no other circuits or nearby antennas influence it.
            In the case of Hertzian dipoles, it is commonly assumed that loading coils are inserted between each antenna and its feedline to compensate for the capacitive reactance~\cite[Ch.~2]{pozar_microwave_2012}.
            Under this assumption, the self-impedance reduces to the \textit{radiation resistance},
            \begin{equation}
                \label{eq:radiation_resistance}
                \R{r} = \frac{2}{3}\pi\eta\left(\frac{l}{\lambda}\right)^2,
            \end{equation}
            with $\eta$ the impedance of free space and $l$ the dipole length.

            Consider now a pair of side-by-side Hertzian dipoles separated by a distance $r$.
            Their mutual impedance is given by
            \begin{equation}
                \label{eq:coupling_side_by_side}
                Z_{||}(r) = \frac{3}{2}\R{r}\im\euler^{-\im \kappa r}\psi(r),
            \end{equation}
            where $\psi(r)$ is an auxiliary function defined as
            \begin{equation}
                \psi(r) = \frac{1}{\kappa r} - \frac{\im}{(\kappa r)^2} - \frac{1}{(\kappa r)^3}
            \end{equation}
            and $\kappa = 2\pi/\lambda$ denotes the wavenumber~\cite{yordanov_arrays_2009}.

            Next, we particularize equations above to obtain the near-field multipath channel.

        \subsection{Multipath Propagation}
            \label{sec:multipath_propagation}
            The mutual impedance expression in~\cref{eq:coupling_side_by_side} is applicable to dipoles belonging to different arrays.
            In fact, evaluating the coupling between a distant source located at $(r,\theta)$ and each antenna element of an array yields the near-field array response vector:
            \begin{equation}
                [\aresp{}(r,\theta)]_n = g_n \frac{\euler^{-\im \kappa r_n}}{\kappa r_n},
            \end{equation}
            where $r_n$ denotes the distance between the source and the $n$-th antenna element, while $g_n\in\cnumbers$ accounts for the element-dependent gain and phase response.

            When multiple propagation paths exist between the transmitter and the receiver, the received signal is given by the superposition of the contributions associated with each path.
            Under a geometric model with $L$ propagation paths, the impedance vector can be expressed as
            \begin{equation}
                \label{eq:z_nlos}
                \bsfz{NLoS} = \sum_{l=1}^{L} \alpha_l \aresp{}(r_l,\theta_l),
            \end{equation}
            where $\alpha_l \in \cnumbers$ is the complex gain of the $l$-th path, and $(r_l,\theta_l)$ denotes the scatterer location.
            As $L\to\infty$, model~\cref{eq:z_nlos} converges to correlated Rayleigh fading $\bsfz{NLoS} \sim \complexnormal(\zeros{\N{}}, \cov{\text{NLoS}})$, where the covariance matrix is given by
            \begin{equation}
                \label{eq:cov_nlos}
                \cov{\text{NLoS}} = \sum_{l=1}^{L} \beta_l \aresp{}(r_l,\theta_l)\herm{\aresp{}(r_l,\theta_l)}
            \end{equation}
            with $\beta_l = \expec{|\alpha_l|^2}$ and $\expec{\alpha_l \alpha_m^*} = 0$, for $l \neq m$.

            Equations~\cref{eq:coupling_side_by_side,eq:z_nlos} will be useful in the sequel for constructing the impedance matrices arising in multiport communication theory, which in turn enable the derivation of physically consistent expressions for the channel $\bsf{h}$.

    \section{Multiport Communication Theory}
        Multiport communication theory provides a circuit-theoretic framework in which the inputs and outputs of a \acrshort{mimo} communication system are associated with the ports of a multiport black box characterized by impedance or scattering matrices~\cite{ivrlac_toward_2010,damico_holographic_2024,mezghani_reincorporating_2024}.
        Within this framework, the complexity of electromagnetic field theory is encapsulated in the circuit model, thereby simplifying the mathematical treatment.
        Moreover, noise modeling is generally more tractable in circuit theory than in electromagnetic field theory, since the latter is inherently deterministic~\cite{ivrlac_toward_2010}.

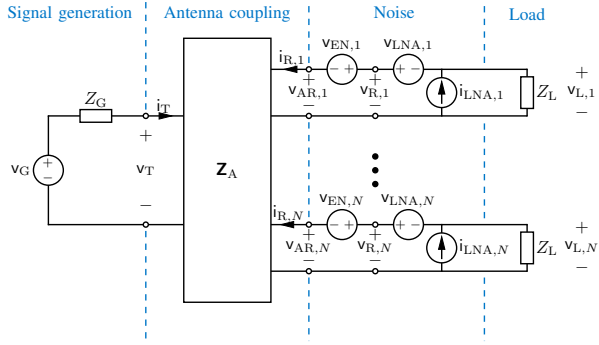
\begin{figure}[t]
            \centering
            \resizebox{0.91\columnwidth}{!}{%
                \begin{tikzpicture}[scale=2.54]%
% dpic version 2025.08.01 option -g for TikZ and PGF 1.01
\ifx\dpiclw\undefined\newdimen\dpiclw\fi
\global\def\dpicdraw{\draw[line width=\dpiclw]}
\global\def\dpicstop{;}
\dpiclw=0.8bp
\dpiclw=0.8bp
\dpicdraw (0,-1.043307) rectangle (0.688976,1.043307)\dpicstop
\draw (0.344488,0) node{$\boldsymbol{\mathsf{Z}}_{\mathrm{A}}$};
\dpicdraw (0,0.429597)
 --(-0.3,0.429597)\dpicstop
\dpicdraw (0,-0.429597)
 --(-0.3,-0.429597)\dpicstop
\dpicdraw[fill=white](-0.3,0.429597) circle (0.007874in)\dpicstop
\dpicdraw[fill=white](-0.3,-0.429597) circle (0.007874in)\dpicstop
\draw (-0.3,0.275591) node{$ +$};
\draw (-0.3,-0) node{$ \mathsf{v}_{\mathrm{T}}$};
\draw (-0.3,-0.275591) node{$ -$};
\dpicdraw (-0.32,0.429597)
 --(-0.57,0.429597)\dpicstop
\dpicdraw (-0.82,0.429597)
 --(-0.82,0.379597)
 --(-0.57,0.379597)
 --(-0.57,0.479597)
 --(-0.82,0.479597)
 --(-0.82,0.429597)\dpicstop
\dpicdraw (-0.82,0.429597)
 --(-1.07,0.429597)\dpicstop
\draw (-0.695,0.479597) node[above=-2bp]{$ Z_{\mathrm{G}}$};
\filldraw[line width=0bp](-0.195,0.404597)
 --(-0.095,0.429597)
 --(-0.195,0.454597) --cycle\dpicstop
\dpicdraw (-0.117906,0.429597)
 --(-0.195,0.429597)\dpicstop
\draw (-0.156453,0.429597) node[above=-2bp]{$ \mathsf{i}_{\mathrm{T}}$};
\dpicdraw (-1.064444,0.429597)
 --(-1.075556,0.429597)\dpicstop
\dpicdraw (-1.07,-0.429597)
 --(-1.07,-0.11811)\dpicstop
\dpicdraw (-1.07,-0) circle (0.0465in)\dpicstop
\draw (-1.07,-0.059055) node{$_-$};
\draw (-1.07,0.059055) node{$_+$};
\dpicdraw (-1.07,0.11811)
 --(-1.07,0.429597)\dpicstop
\draw (-1.18811,0) node[left=-2bp]{$ \mathsf{v}_{\mathrm{G}}$};
\dpicdraw (-1.07,-0.424041)
 --(-1.07,-0.435153)\dpicstop
\dpicdraw (-1.07,-0.429597)
 --(-0.32,-0.429597)\dpicstop
\dpicdraw (0.688976,0.797823)
 --(0.988976,0.797823)\dpicstop
\filldraw[line width=0bp](0.89472,0.822823)
 --(0.79472,0.797823)
 --(0.89472,0.772823) --cycle\dpicstop
\dpicdraw (0.817626,0.797823)
 --(0.89472,0.797823)\dpicstop
\draw (0.856173,0.797823) node[above=-2bp]{$ \mathsf{i}_{\mathrm{R},1}$};
\dpicdraw (0.988976,0.797823)
 --(1.133366,0.797823)\dpicstop
\dpicdraw (1.251476,0.797823) circle (0.0465in)\dpicstop
\draw (1.192421,0.797823) node{$_-$};
\draw (1.310531,0.797823) node{$_+$};
\dpicdraw (1.369587,0.797823)
 --(1.513976,0.797823)\dpicstop
\draw (1.251476,0.915933) node[above=-2bp]{$ \mathsf{v}_{\mathrm{EN},1}$};
\dpicdraw (2.038976,0.797823)
 --(1.894587,0.797823)\dpicstop
\dpicdraw (1.776476,0.797823) circle (0.0465in)\dpicstop
\draw (1.835531,0.797823) node{$_-$};
\draw (1.717421,0.797823) node{$_+$};
\dpicdraw (1.658366,0.797823)
 --(1.513976,0.797823)\dpicstop
\draw (1.776476,0.915933) node[above=-2bp]{$ \mathsf{v}_{\mathrm{LNA},1}$};
\dpicdraw (2.033421,0.797823)
 --(2.044532,0.797823)\dpicstop
\dpicdraw (2.038976,0.429597)
 --(2.038976,0.4956)\dpicstop
\dpicdraw (2.038976,0.61371) circle (0.0465in)\dpicstop
\filldraw[line width=0bp](2.063976,0.602293)
 --(2.038976,0.702293)
 --(2.013976,0.602293) --cycle\dpicstop
\dpicdraw (2.038976,0.525127)
 --(2.038976,0.679387)\dpicstop
\dpicdraw (2.038976,0.73182)
 --(2.038976,0.797823)\dpicstop
\draw (2.157087,0.61371) node[right=-2bp]{$ \mathsf{i}_{\mathrm{LNA},1}$};
\dpicdraw (0.688976,0.429597)
 --(0.988976,0.429597)\dpicstop
\dpicdraw (0.988976,0.429597)
 --(1.513976,0.429597)\dpicstop
\dpicdraw (1.513976,0.429597)
 --(2.038976,0.429597)\dpicstop
\dpicdraw (2.033421,0.429597)
 --(2.044532,0.429597)\dpicstop
\dpicdraw[fill=white](0.988976,0.797823) circle (0.007874in)\dpicstop
\dpicdraw[fill=white](0.988976,0.429597) circle (0.007874in)\dpicstop
\draw (0.988976,0.716072) node{$ +$};
\draw (0.988976,0.61371) node{$ \mathsf{v}_{\mathrm{AR},1}$};
\draw (0.988976,0.511348) node{$ -$};
\dpicdraw[fill=white](1.513976,0.797823) circle (0.007874in)\dpicstop
\dpicdraw[fill=white](1.513976,0.429597) circle (0.007874in)\dpicstop
\draw (1.513976,0.716072) node{$ +$};
\draw (1.513976,0.61371) node{$ \mathsf{v}_{\mathrm{R},1}$};
\draw (1.513976,0.511348) node{$ -$};
\dpicdraw (0.688976,-0.429597)
 --(0.988976,-0.429597)\dpicstop
\filldraw[line width=0bp](0.868884,-0.404597)
 --(0.768884,-0.429597)
 --(0.868884,-0.454597) --cycle\dpicstop
\dpicdraw (0.79179,-0.429597)
 --(0.868884,-0.429597)\dpicstop
\draw (0.830337,-0.429597) node[above=-2bp]{$ \mathsf{i}_{\mathrm{R},N}$};
\dpicdraw (0.988976,-0.429597)
 --(1.133366,-0.429597)\dpicstop
\dpicdraw (1.251476,-0.429597) circle (0.0465in)\dpicstop
\draw (1.192421,-0.429597) node{$_-$};
\draw (1.310531,-0.429597) node{$_+$};
\dpicdraw (1.369587,-0.429597)
 --(1.513976,-0.429597)\dpicstop
\draw (1.251476,-0.311487) node[above=-2bp]{$ \mathsf{v}_{\mathrm{EN},N}$};
\dpicdraw (2.038976,-0.429597)
 --(1.894587,-0.429597)\dpicstop
\dpicdraw (1.776476,-0.429597) circle (0.0465in)\dpicstop
\draw (1.835531,-0.429597) node{$_-$};
\draw (1.717421,-0.429597) node{$_+$};
\dpicdraw (1.658366,-0.429597)
 --(1.513976,-0.429597)\dpicstop
\draw (1.776476,-0.311487) node[above=-2bp]{$ \mathsf{v}_{\mathrm{LNA},N}$};
\dpicdraw (2.033421,-0.429597)
 --(2.044532,-0.429597)\dpicstop
\dpicdraw (2.038976,-0.797823)
 --(2.038976,-0.73182)\dpicstop
\dpicdraw (2.038976,-0.61371) circle (0.0465in)\dpicstop
\filldraw[line width=0bp](2.063976,-0.625127)
 --(2.038976,-0.525127)
 --(2.013976,-0.625127) --cycle\dpicstop
\dpicdraw (2.038976,-0.702293)
 --(2.038976,-0.548034)\dpicstop
\dpicdraw (2.038976,-0.4956)
 --(2.038976,-0.429597)\dpicstop
\draw (2.157087,-0.61371) node[right=-2bp]{$ \mathsf{i}_{\mathrm{LNA},N}$};
\dpicdraw (0.688976,-0.797823)
 --(0.988976,-0.797823)\dpicstop
\dpicdraw (0.988976,-0.797823)
 --(1.513976,-0.797823)\dpicstop
\dpicdraw (1.513976,-0.797823)
 --(2.038976,-0.797823)\dpicstop
\dpicdraw (2.033421,-0.797823)
 --(2.044532,-0.797823)\dpicstop
\dpicdraw[fill=white](0.988976,-0.429597) circle (0.007874in)\dpicstop
\dpicdraw[fill=white](0.988976,-0.797823) circle (0.007874in)\dpicstop
\draw (0.988976,-0.511348) node{$ +$};
\draw (0.988976,-0.61371) node{$ \mathsf{v}_{\mathrm{AR},N}$};
\draw (0.988976,-0.716072) node{$ -$};
\dpicdraw[fill=white](1.513976,-0.429597) circle (0.007874in)\dpicstop
\dpicdraw[fill=white](1.513976,-0.797823) circle (0.007874in)\dpicstop
\draw (1.513976,-0.511348) node{$ +$};
\draw (1.513976,-0.61371) node{$ \mathsf{v}_{\mathrm{R},N}$};
\draw (1.513976,-0.716072) node{$ -$};
\dpicdraw[fill=black](1.513976,0.109113) circle (0.007874in)\dpicstop
\dpicdraw[fill=black](1.513976,-0) circle (0.007874in)\dpicstop
\dpicdraw[fill=black](1.513976,-0.109113) circle (0.007874in)\dpicstop
\dpicdraw (2.038976,0.797823)
 --(2.713976,0.797823)\dpicstop
\dpicdraw (2.708421,0.797823)
 --(2.719532,0.797823)\dpicstop
\dpicdraw (2.713976,0.797823)
 --(2.713976,0.73871)\dpicstop
\dpicdraw (2.713976,0.48871)
 --(2.763976,0.48871)
 --(2.763976,0.73871)
 --(2.663976,0.73871)
 --(2.663976,0.48871)
 --(2.713976,0.48871)\dpicstop
\dpicdraw (2.713976,0.48871)
 --(2.713976,0.429597)\dpicstop
\draw (2.763976,0.61371) node[right=-2bp]{$ Z_{\mathrm{L}}$};
\draw (3.147047,0.77119) node{$ +$};
\draw (3.147047,0.61371) node{$ \mathsf{v}_{\mathrm{L},1}$};
\draw (3.147047,0.45623) node{$ -$};
\dpicdraw (2.713976,0.435153)
 --(2.713976,0.424041)\dpicstop
\dpicdraw (2.713976,0.429597)
 --(2.038976,0.429597)\dpicstop
\dpicdraw (2.038976,-0.429597)
 --(2.713976,-0.429597)\dpicstop
\dpicdraw (2.708421,-0.429597)
 --(2.719532,-0.429597)\dpicstop
\dpicdraw (2.713976,-0.429597)
 --(2.713976,-0.48871)\dpicstop
\dpicdraw (2.713976,-0.73871)
 --(2.763976,-0.73871)
 --(2.763976,-0.48871)
 --(2.663976,-0.48871)
 --(2.663976,-0.73871)
 --(2.713976,-0.73871)\dpicstop
\dpicdraw (2.713976,-0.73871)
 --(2.713976,-0.797823)\dpicstop
\draw (2.763976,-0.61371) node[right=-2bp]{$ Z_{\mathrm{L}}$};
\draw (3.147047,-0.45623) node{$ +$};
\draw (3.147047,-0.61371) node{$ \mathsf{v}_{\mathrm{L},N}$};
\draw (3.147047,-0.77119) node{$ -$};
\dpicdraw (2.713976,-0.792268)
 --(2.713976,-0.803379)\dpicstop
\dpicdraw (2.713976,-0.797823)
 --(2.038976,-0.797823)\dpicstop
\definecolor{lcspec}{rgb}{0.00000,0.46700,0.78431}%
\color[rgb]{0.00000,0.46700,0.78431}%
\global\let\dpiclidraw=\dpicdraw\global\let\dpicfidraw=\filldraw
\global\def\dpicdraw{\dpiclidraw[color=lcspec]}
\global\def\filldraw{\dpicfidraw[color=lcspec]}
\dpicdraw[dash pattern=on 0.05in off 0.05in](-0.3,0.479597)
 --(-0.3,1.35)\dpicstop
\dpicdraw[dash pattern=on 0.05in off 0.05in](-0.3,-0.479597)
 --(-0.3,-1.35)\dpicstop
\dpicdraw[dash pattern=on 0.05in off 0.05in](0.988976,0.847823)
 --(0.988976,1.35)\dpicstop
\dpicdraw[dash pattern=on 0.05in off 0.05in](0.988976,-0.847823)
 --(0.988976,-1.35)\dpicstop
\dpicdraw[dash pattern=on 0.05in off 0.05in](0.988976,0.379597)
 --(0.988976,-0.379597)\dpicstop
\dpicdraw[dash pattern=on 0.05in off 0.05in](2.376476,0.847823)
 --(2.376476,1.35)\dpicstop
\dpicdraw[dash pattern=on 0.05in off 0.05in](2.376476,-0.847823)
 --(2.376476,-1.35)\dpicstop
\dpicdraw[dash pattern=on 0.05in off 0.05in](2.376476,-0.379597)
 --(2.376476,0.379597)\dpicstop
\draw (-0.9,1.2375) node(Text){Signal generation};
\draw (0.344488,1.2375) node{Antenna coupling};
\draw (1.663976,1.2475) node{Noise};
\draw (2.713976,1.2375) node{Load};
\end{tikzpicture}%
            }
            \caption{Multiport communication model of a \acrshort{simo} system.}
            \label{fig:multiport_simo}
        \end{figure}

        The multiport model for the particular case of \acrshort{simo} is illustrated in~\cref{fig:multiport_simo}.
        It consists of four main components: signal generation, impedance matching, antenna mutual coupling, and noise.
        Impedance matching is not illustrated in \cref{fig:multiport_simo} since it is omitted in the considered scenario, as we shall discuss next.
        Furthermore, since we are interested in propagation-related aspects, we focus primarily on antenna mutual coupling~(see~\cite{ivrlac_toward_2010},~\cite{damico_holographic_2024} for further details).

        \subsection{Impedance Matching}
            Impedance matching networks can be beneficial when placed between the antenna array and the amplifiers or signal generators.
            These networks can be designed to achieve specific objectives, such as maximizing power transfer from the signal generators to the antennas (\ie power matching) or optimizing the \acrlong{snr} at the outputs of the receive amplifiers (\ie noise matching).
            In general, matching networks are assumed to be lossless, reciprocal and noiseless~\cite[Sec.~4.2]{pozar_microwave_2012}.

            However, when the number of antennas becomes very large, implementing optimal matching networks is highly challenging~\cite{ivrlac_toward_2010,mezghani_reincorporating_2024}, so they are often replaced by self-impedance matching networks~\cite{damico_holographic_2024,warnick_minimizing_2009}.
            Moreover, if the unilateral approximation does not hold, the matching network expressions become coupled transcendental equations with no closed-form solution~\cite{mezghani_reincorporating_2024,phang_near-field_2018,marti_asymptotic_2024}.

            Since this work investigates the validity of the unilateral approximation in large arrays, matching networks are not considered.
            Furthermore, the design and implementation of matching networks constitute a complex problem in their own right, so this choice isolates the impact of receiver-to-transmitter coupling from that of impedance matching.

        \subsection{Antenna Mutual Coupling}
            Antenna mutual coupling can be divided into four components: intra-array coupling at the transmitter, intra-array coupling at the receiver, inter-array coupling from the transmitter to the receiver (\ie the communication channel itself), and inter-array coupling from the receiver to the transmitter.
            In this work, we are primarily interested in characterizing the latter component.

            Within the multiport communication framework, antenna mutual coupling is modeled through the impedance network $\bsfZ{A}$, which can be partitioned according to the four components described above:
            \begin{equation}
                \label{eq:z_a}
                \bsfZ{A} = \begin{pmatrix}
                    \Z{T} & \bsfz[T]{TR}\\
                    \bsfz{RT} & \bfZ{R}
                \end{pmatrix} \in \cnumbers^{(N+1)\times(N+1)},
            \end{equation}
            where $\Z{T}$ denotes the transmitting antenna self-impedance, $\bfZ{R}$ the receiving-array coupling matrix, and $\bsfz{RT}$ and $\bsfz{TR}$ the transmitter-to-receiver and receiver-to-transmitter coupling vectors.
            Since antennas are reciprocal devices~\cite[Sec.~3.8]{balanis_antenna_2016}, it is satisfied that $\bsfZ{A} = \bsfZ[T]{A}$.
            Moreover, we assume lossless antennas, so $\Re(\bsfZ{A}) = 0.$

            Each block in \cref{eq:z_a} is modeled according to the physical characteristics of the employed antennas and the propagation environment.
            In our case, we will employ the mutual impedance expressions derived in~\cref{sec:coupling_dipoles,sec:multipath_propagation}.
            In particular, inter-array coupling is given by $\bsfz{RT} = \bsfz{TR} = \bsfz{NLoS}$ in \cref{eq:z_nlos} and intra-array coupling by
            \begin{equation}
                \label{eq:z_r_def}
                \begin{cases}
                    [\bfZ{R}]_{m,n} = \R{r},\ m=n,\ 1\leq m, n \leq N\\
                    [\bfZ{R}]_{m,n} = Z_{||}(d |m-n|),\ m \neq n,
                \end{cases}
            \end{equation}
            with $\R{r}$ defined in \cref{eq:radiation_resistance} and $Z_{||}$ in \cref{eq:coupling_side_by_side}.

        \subsection{Input-Output Relation}
            Multiport systems such as the one considered herein are linear.
            Therefore, the input-output relation must be of the form
            \begin{equation}
                \label{eq:multiport_in_out}
                \bsfvs{L} = \bsf{d}\sfvs{G} + \bsf{n},
            \end{equation}
            where we define $\bsfvs{L} = (\sfvs{L,1},\dots,\mathsf{v}_{\mathrm{L},N})$, and analogously for the remaining voltages and currents appearing in~\cref{fig:multiport_simo}.

            In \cref{eq:multiport_in_out}, $\bsf{d}\sfvs{G}$ denotes the noise-free component and $\bsf{n}$ the signal-free component.
            Since our main interest lies in characterizing the channel vector, we will focus exclusively on the former.

            From the definition of impedance parameters~\cite[Sec.~4.2]{pozar_microwave_2012}, the antenna multiport voltages can be expressed in terms of the currents as
            \begin{equation}
                \begin{pmatrix}
                    \sfvs{T}\\
                    \bsfvs{R}
                \end{pmatrix} = \begin{pmatrix}
                    \Z{T} & \bsfz[T]{TR}\\
                    \bsfz{RT} & \bfZ{R}
                \end{pmatrix} \begin{pmatrix}
                    \sfis{T}\\
                    \bsfis{R}
                \end{pmatrix}.
            \end{equation}
            Furthermore, from Ohm's law, $\sfvs{T} = \sfvs{G} - \Z{G}\sfis{T}$ and, since all noise sources are set to zero, $\bsfvs{L} = \bsfvs{R}$.
            After some manipulations, the physical uplink channel is given by
            \begin{equation}
                \label{eq:physical_channel}
                \bsf{d} = \frac{\Z{L}}{\Z{G}+\Z{T}}\left(\Z{L}\id{N}+\bfZ{R}-\frac{\bsfz{RT}\bsfz[T]{TR}}{\Z{G}+\Z{T}}\right)^{-1}\bsfz{RT},
            \end{equation}
            which is related to the information-theoretic channel $\bsf{h}$ through a linear transformation~\cite{damico_holographic_2024}.
            Consequently, if $\bsf{d}$ follows a complex normal distribution, then $\bsf{h}$ does as well.\footnote{The linear transformation relating $\bsf{d}$ and $\bsf{h}$ is deterministic under the unilateral approximation.}

        \subsection{Unilateral Approximation}
            In many practical scenarios, it is reasonable to assume that the currents at the receiving antennas have a negligible effect on the transmitter, mainly due to the large separation between them~\cite{ivrlac_toward_2010}.
            Under this assumption, the antenna coupling network reduces to
            \begin{equation}
                \bsfZ[UA]{A} = \begin{pmatrix}
                    \Z{T} & \zeros{}\\
                    \bsfz{RT} & \bfZ{R}
                \end{pmatrix},
            \end{equation}
            which corresponds to the impedance matrix of a unilateral network~\cite[Sec.~11.2]{pozar_microwave_2012}.
            For this reason, $\bsfz{TR}\approx\zeros{}$ is referred to as the \textit{unilateral approximation}.

            By setting $\bsfz{TR}=\zeros{}$ in \cref{eq:physical_channel}, we obtain
            \begin{equation}
                \label{eq:physical_channel_ua}
                \bsf{d}_{\text{UA}} = \frac{\Z{L}}{\Z{G}+\Z{T}}\left(\Z{L}\id{N}+\bfZ{R}\right)^{-1}\bsfz{RT}.
            \end{equation}
            Thus, under the unilateral approximation, the communication channel follows the same distribution as $\bsfz{RT}$, up to a deterministic linear transformation.

            Comparing~\cref{eq:physical_channel} and~\cref{eq:physical_channel_ua}, we observe that a finer approximation is given by
            \begin{equation}
                \label{eq:unilateral_approximation}
                \norm{\frac{\bsfz{RT}\bsfz[T]{TR}}{\Z{G}+\Z{T}}}{F} \ll \norm{\Z{L}\id{N}+\bfZ{R}}{F},
            \end{equation}
            so $\bsf{d} \approx \bsf[UA]{d}$.

            The validity of \cref{eq:unilateral_approximation} in the near field has so far only been established under \acrshort{los} conditions~\cite{marti_asymptotic_2024}.
            In order to extend it to \acrshort{nlos} propagation, the analysis must account for the inherent stochastic properties of the mobile channel.
            Hence, we will assess \cref{eq:unilateral_approximation} on average.
            Formally,
            \begin{equation}
                \label{eq:ua_nlos}
                \frac{\expec{\norm{\bsfz{RT}\bsfz[T]{TR}}{F}}}{|\Z{G}+\Z{T}|} \ll \norm{\Z{L}\id{N}+\bfZ{R}}{F},
            \end{equation}
            which follows from the fact that $\bsfz{RT}$ and $\bsfz{TR}$ are the only random quantities in~\cref{eq:unilateral_approximation}.

            Taking into account that $\bsfz{RT}\bsfz[T]{TR}$ is a rank-one matrix, it follows that $\norm{\bsfz{RT}\bsfz[T]{TR}}{F} = \norm{\bsfz{RT}}{2}^2$.
            Thus, condition~\cref{eq:ua_nlos} can be rewritten as
            \begin{equation}
                \label{eq:inter_array_coupling}
                \frac{\trace{\cov{\mathrm{RT}}}}{|\Z{G}+\Z{T}|} \ll \norm{\Z{L}\id{N}+\bfZ{R}}{F}.
            \end{equation}
            In our case, we will set $\cov{\mathrm{RT}}=\cov{\text{NLoS}}$ as defined in \cref{eq:cov_nlos}.

    \section{Asymptotic Analysis}
        In this section, we investigate the conditions under which~\cref{eq:inter_array_coupling} holds.
        In particular, we focus on the large-array regime, \ie $N\to\infty$.
        Two scenarios are considered: discrete arrays with fixed inter-element spacing $d$, for which $N\to\infty$ implies $D\to\infty$, and continuous arrays with fixed aperture $D$, for which $N\to\infty$ implies $d\to0$.

        Using the model described in~\cref{eq:z_r_def}, the right-hand side of~\cref{eq:inter_array_coupling} can be lower bounded as
        \begin{equation}
            \label{eq:bound_intra}
            \begin{aligned}
                &\norm{\Z{L}\id{N}+\bfZ{R}}{F}^2 \geq N|\Z{L}+\R{r}|^2 \\
                &\qquad+\frac{9}{4}N\R{r}^2\bigg(\frac{\harmn{N-1}{2}}{(\kappa d)^2}-\frac{\harmn{N-1}{4}}{(\kappa d)^4} + \frac{\harmn{N-1}{6}}{(\kappa d)^6}\bigg),
            \end{aligned}
        \end{equation}
        where $\harmn{k}{q}$ denotes the generalized harmonic number of order $q$~\cite[Sec.~1.2.7]{knuth_art_1997}.
        This bound is obtained by multiplying by $N$ the norm of a row corresponding to an edge antenna, since edge antennas exhibit the weakest coupling (see~\cite{marti_asymptotic_2024} for further details).

        On the other hand, the left-hand side is given by
        \begin{equation}
            \label{eq:trace_cov}
            \frac{\trace{\cov{\mathrm{RT}}}}{|\Z{G}+\Z{T}|} = \frac{1}{|\Z{G}+\Z{T}|}\sum_{l=1}^{L}\beta_l\norm{\aresp{}(r_l,\theta_l)}{2}^2.
        \end{equation}
        As in \cite{marti_asymptotic_2024}, we assume $\theta_l = \pi/2$ so the (squared) distance from the $l$-th scatterer to the $n$-th antenna in the array is
        \begin{equation}
            r_{l,n}^2 = r_l^2+(nd)^2,\quad -\frac{N-1}{2} \leq n \leq \frac{N-1}{2}.
        \end{equation}
        Thus, the norm of the array response vector is given by
        \begin{equation}
            \label{eq:ar_ula}
            \norm{\aresp{}\left(r_l,\pi/2\right)}{2}^2 = \frac{\R{r}^2}{\kappa^2}\left(2\!\sum_{n=0}^{\frac{N-1}{2}}\frac{1}{r_l^2+(nd)^2} - \frac{1}{r_l^2}\right),
        \end{equation}
        up to an angle-dependent scaling factor (\ie $g_n$).

        Next, we particularize expressions \cref{eq:bound_intra,eq:ar_ula} for discrete and continuous arrays.

        \subsection{Discrete Arrays}
            Assume that $d$ is a constant, so $D\to\infty$ when $N\to\infty$.
            Then, \cref{eq:bound_intra} diverges linearly as $N\to\infty$.
            Since \cref{eq:bound_intra} is the square of the intra-array coupling norm, it follows that
            \begin{equation}
                \norm{\Z{L}\id{N}+\bfZ{R}}{F} = O(\sqrt{N}),
            \end{equation}
            as previously proved in \cite{marti_asymptotic_2024}.

            Regarding inter-array coupling, the limit when $N\to\infty$ of~\cref{eq:ar_ula} converges and can be computed in closed-form using the Poisson summation formula:
            \begin{equation}
                \label{eq:limit_inter}
                \lim_{N\to\infty} \norm{\aresp{}\left(r_l,\pi/2\right)}{2}^2 = \frac{\R{r}^2 \pi r_l \coth\left(\frac{\pi r_l}{d}\right)}{d\kappa^2 r_l^2} < +\infty.
            \end{equation}
            Since the remaining terms in \cref{eq:trace_cov} do not depend on $N$, the left-hand side in \cref{eq:inter_array_coupling} is bounded when $N\to\infty$.
            This same behavior was observed in \acrshort{los} propagation.

            Therefore, \cref{eq:inter_array_coupling} holds asymptotically for discrete arrays.
            In~\cref{sec:numerical_results}, we illustrate the behavior of mutual coupling for a finite number of antennas, as well as for the case $\theta_l \neq \pi/2$.

        \subsection{Continuous Arrays}
            Now consider a \acrshort{ula} with fixed aperture so $d=D/(N-1)$.
            Following a procedure analogous to that in~\cite{marti_asymptotic_2024}, it follows that
            \begin{equation}
                \norm{\Z{L}\id{N}+\bfZ{R}}{F} = O(N^{3.5}).
            \end{equation}

            Similarly, substituting $d$ into \cref{eq:ar_ula} and bounding each series term with $\frac{1}{r_l^2+D^2}$ yields a linear divergence.
            That is,
            \begin{equation}
                \frac{\trace{\cov{\mathrm{RT}}}}{|\Z{G}+\Z{T}|} = O(N).
            \end{equation}
            Once again, this corresponds to the same divergence rate obtained under \acrshort{los} conditions~\cite{marti_asymptotic_2024}.

            Since the divergence order of inter-array coupling is smaller than that of intra-array coupling, we conclude that~\cref{eq:inter_array_coupling} is asymptotically satisfied.
            Therefore, the unilateral approximation holds on average for sufficiently large $N$.

            This result provides a theoretical justification for simplifying the coupling model in large-scale antenna systems, thereby enabling the use of correlated Rayleigh fading models for the information-theoretic channel.

    \section{Numerical Results}
        \label{sec:numerical_results}
        To validate the theoretical results derived in the previous section, we now consider two scenarios employing realistic parameters.
        These simulations illustrate the behavior of mutual coupling in the non-asymptotic regime, as well as in the case where scatterers are arbitrarily located (\ie not necessarily orthogonal to the array).

        The first scenario corresponds to discrete arrays with $d=\lambda/2$, whereas the second considers a continuous array with fixed aperture $D=\qty{5}{\m}$.
        In both cases, we consider a \acrshort{ue} surrounded by $L$ uniformly distributed scatterers within a sphere of radius $r_{\mathrm{s}}=\qty{3}{\m}$.
        The carrier frequency is set to $f=\qty{10}{\GHz}$, the dipole length to $l=\lambda/20$, and the load and generator impedances to $\Z{G}=\Z{L}=186-\im31.6\,\unit{\ohm}$, as measured in~\cite{lehmeyer_lna_2015}.
        Two different \acrshort{ue} locations and scattering environments are considered in each simulation.
        The simulation parameters are summarized in~\cref{tab:simulation_parameters}.
        \begin{table}
            \renewcommand{\arraystretch}{1.4}
            \centering
            \caption{Summary of simulation parameters.}
            \begin{tabular}{ll}
                \toprule
                \textbf{Parameter} & \textbf{Value} \\
                \midrule
                Carrier frequency              & $f = \qty{10}{\GHz}$ \\
                Amplifier and load impedance   & $\Z{G} = \Z{L} = 186 - \im 31.6\,\unit{\ohm}$ \\
                Dipole length                  & $l = \lambda / 20$ \\
                Scattering sphere radius       & $r_{\mathrm{s}} = \qty{3}{\meter}$ \\
                \bottomrule
            \end{tabular}
            \label{tab:simulation_parameters}
        \end{table}

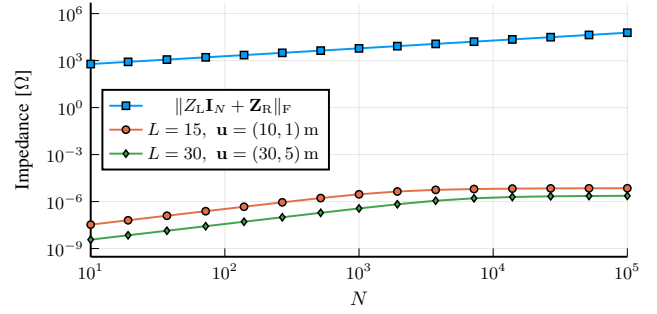
\begin{figure}
            \centering
            \resizebox{\columnwidth}{!}{%
                % Recommended preamble:
% \usetikzlibrary{arrows.meta}
% \usetikzlibrary{backgrounds}
% \usepgfplotslibrary{patchplots}
% \usepgfplotslibrary{fillbetween}
% \pgfplotsset{%
%     layers/standard/.define layer set={%
%         background,axis background,axis grid,axis ticks,axis lines,axis tick labels,pre main,main,axis descriptions,axis foreground%
%     }{
%         grid style={/pgfplots/on layer=axis grid},%
%         tick style={/pgfplots/on layer=axis ticks},%
%         axis line style={/pgfplots/on layer=axis lines},%
%         label style={/pgfplots/on layer=axis descriptions},%
%         legend style={/pgfplots/on layer=axis descriptions},%
%         title style={/pgfplots/on layer=axis descriptions},%
%         colorbar style={/pgfplots/on layer=axis descriptions},%
%         ticklabel style={/pgfplots/on layer=axis tick labels},%
%         axis background@ style={/pgfplots/on layer=axis background},%
%         3d box foreground style={/pgfplots/on layer=axis foreground},%
%     },
% }

\begin{tikzpicture}[/tikz/background rectangle/.style={fill={rgb,1:red,1.0;green,1.0;blue,1.0}, fill opacity={1.0}, draw opacity={1.0}}, show background rectangle]
\begin{axis}[point meta max={nan}, point meta min={nan}, legend cell align={left}, legend columns={1}, title={}, title style={at={{(0.5,1)}}, anchor={south}, font={{\fontsize{14 pt}{18.2 pt}\selectfont}}, color={rgb,1:red,0.0;green,0.0;blue,0.0}, draw opacity={1.0}, rotate={0.0}, align={center}}, legend style={color={rgb,1:red,0.0;green,0.0;blue,0.0}, draw opacity={1.0}, line width={1}, solid, fill={rgb,1:red,1.0;green,1.0;blue,1.0}, fill opacity={1.0}, text opacity={1.0}, font={{\fontsize{9 pt}{11.700000000000001 pt}\selectfont}}, text={rgb,1:red,0.0;green,0.0;blue,0.0}, cells={anchor={center}}, at={(0.02, 0.5)}, anchor={west}}, axis background/.style={fill={rgb,1:red,1.0;green,1.0;blue,1.0}, opacity={1.0}}, anchor={north west}, xshift={1.0mm}, yshift={-1.0mm}, width={112.3mm}, height={61.5mm}, scaled x ticks={false}, xlabel={$N$}, x tick style={color={rgb,1:red,0.0;green,0.0;blue,0.0}, opacity={1.0}}, x tick label style={color={rgb,1:red,0.0;green,0.0;blue,0.0}, opacity={1.0}, rotate={0}}, xlabel style={at={(ticklabel cs:0.5)}, anchor=near ticklabel, at={{(ticklabel cs:0.5)}}, anchor={near ticklabel}, font={{\fontsize{10 pt}{13.0 pt}\selectfont}}, color={rgb,1:red,0.0;green,0.0;blue,0.0}, draw opacity={1.0}, rotate={0.0}}, xmode={log}, log basis x={10}, xmajorgrids={true}, xmin={10.0}, xmax={100000.0}, xticklabels={{$10^{1}$,$10^{2}$,$10^{3}$,$10^{4}$,$10^{5}$}}, xtick={{10.0,100.0,1000.0,10000.0,100000.0}}, xtick align={inside}, xticklabel style={font={{\fontsize{9 pt}{11.700000000000001 pt}\selectfont}}, color={rgb,1:red,0.0;green,0.0;blue,0.0}, draw opacity={1.0}, rotate={0.0}}, x grid style={color={rgb,1:red,0.0;green,0.0;blue,0.0}, draw opacity={0.1}, line width={0.5}, solid}, axis x line*={left}, x axis line style={color={rgb,1:red,0.0;green,0.0;blue,0.0}, draw opacity={1.0}, line width={1}, solid}, scaled y ticks={false}, ylabel={Impedance [$\Omega$]}, y tick style={color={rgb,1:red,0.0;green,0.0;blue,0.0}, opacity={1.0}}, y tick label style={color={rgb,1:red,0.0;green,0.0;blue,0.0}, opacity={1.0}, rotate={0}}, ylabel style={at={(ticklabel cs:0.5)}, anchor=near ticklabel, at={{(ticklabel cs:0.5)}}, anchor={near ticklabel}, font={{\fontsize{10 pt}{13.0 pt}\selectfont}}, color={rgb,1:red,0.0;green,0.0;blue,0.0}, draw opacity={1.0}, rotate={0.0}}, ymode={log}, log basis y={10}, ymajorgrids={true}, ymin={3.0e-10}, ymax={5.0e6}, yticklabels={{$10^{-9}$,$10^{-6}$,$10^{-3}$,$10^{0}$,$10^{3}$,$10^{6}$}}, ytick={{1.0e-9,1.0e-6,0.001,1.0,1000.0,1.0e6}}, ytick align={inside}, yticklabel style={font={{\fontsize{9 pt}{11.700000000000001 pt}\selectfont}}, color={rgb,1:red,0.0;green,0.0;blue,0.0}, draw opacity={1.0}, rotate={0.0}}, y grid style={color={rgb,1:red,0.0;green,0.0;blue,0.0}, draw opacity={0.1}, line width={0.5}, solid}, axis y line*={left}, y axis line style={color={rgb,1:red,0.0;green,0.0;blue,0.0}, draw opacity={1.0}, line width={1}, solid}, colorbar={false}]
    \addplot[color={rgb,1:red,0.0;green,0.6056;blue,0.9787}, name path={4}, draw opacity={1.0}, line width={1}, solid, mark={square*}, mark size={1.875 pt}, mark repeat={1}, mark options={color={rgb,1:red,0.0;green,0.0;blue,0.0}, draw opacity={1.0}, fill={rgb,1:red,0.0;green,0.6056;blue,0.9787}, fill opacity={1.0}, line width={0.75}, rotate={0}, solid}]
        table[row sep={\\}]
        {
            \\
            10.0  602.7686071492805  \\
            19.0  830.8596945686791  \\
            37.0  1159.4496582971526  \\
            72.0  1617.3996950175285  \\
            139.0  2247.2870356211824  \\
            268.0  3120.4595892847947  \\
            518.0  4338.264786550979  \\
            1000.0  6027.693779519431  \\
            1931.0  8376.109007505396  \\
            3728.0  11638.288991813126  \\
            7197.0  16170.629892020741  \\
            13895.0  22468.830179222765  \\
            26827.0  31220.312160446236  \\
            51795.0  43380.55513013917  \\
            100000.0  60276.93856698577  \\
        }
        ;
    \addlegendentry {$\|Z_{\mathrm{L}}\mathbf{I}_N+\mathbf{Z}_{\mathrm{R}}\|_{\mathrm{F}}$}
    \addplot[color={rgb,1:red,0.8889;green,0.4356;blue,0.2781}, name path={5}, draw opacity={1.0}, line width={1}, solid, mark={*}, mark size={1.875 pt}, mark repeat={1}, mark options={color={rgb,1:red,0.0;green,0.0;blue,0.0}, draw opacity={1.0}, fill={rgb,1:red,0.8889;green,0.4356;blue,0.2781}, fill opacity={1.0}, line width={0.75}, rotate={0}, solid}]
        table[row sep={\\}]
        {
            \\
            10.0  3.368314872334527e-8  \\
            19.0  6.399510822242891e-8  \\
            37.0  1.2460044264461813e-7  \\
            72.0  2.423067774257411e-7  \\
            139.0  4.666550858744863e-7  \\
            268.0  8.917786818136016e-7  \\
            518.0  1.6697312181726302e-6  \\
            1000.0  2.9127083412902017e-6  \\
            1931.0  4.391228667417645e-6  \\
            3728.0  5.581716752478732e-6  \\
            7197.0  6.310234323653848e-6  \\
            13895.0  6.707270097659459e-6  \\
            26827.0  6.915848010799512e-6  \\
            51795.0  7.0242959552442806e-6  \\
            100000.0  7.080523548903516e-6  \\
        }
        ;
    \addlegendentry {$L=15,\ \mathbf{u}=(10,1)\,\mathrm{m}$}
    \addplot[color={rgb,1:red,0.2422;green,0.6433;blue,0.3044}, name path={6}, draw opacity={1.0}, line width={1}, solid, mark={diamond*}, mark size={1.875 pt}, mark repeat={1}, mark options={color={rgb,1:red,0.0;green,0.0;blue,0.0}, draw opacity={1.0}, fill={rgb,1:red,0.2422;green,0.6433;blue,0.3044}, fill opacity={1.0}, line width={0.75}, rotate={0}, solid}]
        table[row sep={\\}]
        {
            \\
            10.0  3.7067547601046106e-9  \\
            19.0  7.04280096490134e-9  \\
            37.0  1.3714679418262531e-8  \\
            72.0  2.6686192848640956e-8  \\
            139.0  5.150607849683549e-8  \\
            268.0  9.921305475329815e-8  \\
            518.0  1.9108980396441684e-7  \\
            1000.0  3.6418285580955397e-7  \\
            1931.0  6.722122084324208e-7  \\
            3728.0  1.1318455215679919e-6  \\
            7197.0  1.6147761579434952e-6  \\
            13895.0  1.9618783860176386e-6  \\
            26827.0  2.162613234588157e-6  \\
            51795.0  2.26993312948571e-6  \\
            100000.0  2.326002508711224e-6  \\
        }
        ;
    \addlegendentry {$L=30,\ \mathbf{u}=(30,5)\,\mathrm{m}$}
\end{axis}
\end{tikzpicture}
            }
            \caption{Intra-array and inter-array coupling terms in~\cref{eq:inter_array_coupling} as functions of $N$ for fixed inter-element spacing $d=\lambda/2$, under two different propagation scenarios.}
            \label{fig:nlos_ula_spacing}
        \end{figure}

        In \cref{fig:nlos_ula_spacing} we depict $\norm{\Z{L}\id{N}+\bfZ{R}}{F}$ and $\trace{\cov{\mathrm{RT}}}/|\Z{G}+\Z{T}|$ for a fixed inter-element spacing.
        We observe that, as predicted, intra-array coupling grows proportionally to $\sqrt{N}$.
        Moreover, this growth is monotonic for all $N$.
        In contrast, inter-array coupling increases up to a certain point, after which it converges to a weighted sum of~\cref{eq:limit_inter}, with the weights given by the scatterer powers.
        Regarding the two scattering environments considered, when $L=15$ and the \acrshort{ue} is located at $(10, 1)\,\unit{m}$, inter-array coupling converges more slowly and attains larger values, compared to the case with $L=30$ and $\mathbf{u}=(30,5)\,\unit{m}$.
        This behavior is consistent with the presence of more propagation paths, which slow convergence (sum over $L$), and with the reduced distance to the array, which increases coupling.

\begin{figure}
            \centering
            \resizebox{\columnwidth}{!}{%
                % Recommended preamble:
% \usetikzlibrary{arrows.meta}
% \usetikzlibrary{backgrounds}
% \usepgfplotslibrary{patchplots}
% \usepgfplotslibrary{fillbetween}
% \pgfplotsset{%
%     layers/standard/.define layer set={%
%         background,axis background,axis grid,axis ticks,axis lines,axis tick labels,pre main,main,axis descriptions,axis foreground%
%     }{
%         grid style={/pgfplots/on layer=axis grid},%
%         tick style={/pgfplots/on layer=axis ticks},%
%         axis line style={/pgfplots/on layer=axis lines},%
%         label style={/pgfplots/on layer=axis descriptions},%
%         legend style={/pgfplots/on layer=axis descriptions},%
%         title style={/pgfplots/on layer=axis descriptions},%
%         colorbar style={/pgfplots/on layer=axis descriptions},%
%         ticklabel style={/pgfplots/on layer=axis tick labels},%
%         axis background@ style={/pgfplots/on layer=axis background},%
%         3d box foreground style={/pgfplots/on layer=axis foreground},%
%     },
% }

\begin{tikzpicture}[/tikz/background rectangle/.style={fill={rgb,1:red,1.0;green,1.0;blue,1.0}, fill opacity={1.0}, draw opacity={1.0}}, show background rectangle]
\begin{axis}[point meta max={nan}, point meta min={nan}, legend cell align={left}, legend columns={1}, title={}, title style={at={{(0.5,1)}}, anchor={south}, font={{\fontsize{14 pt}{18.2 pt}\selectfont}}, color={rgb,1:red,0.0;green,0.0;blue,0.0}, draw opacity={1.0}, rotate={0.0}, align={center}}, legend style={color={rgb,1:red,0.0;green,0.0;blue,0.0}, draw opacity={1.0}, line width={1}, solid, fill={rgb,1:red,1.0;green,1.0;blue,1.0}, fill opacity={1.0}, text opacity={1.0}, font={{\fontsize{9 pt}{11.700000000000001 pt}\selectfont}}, text={rgb,1:red,0.0;green,0.0;blue,0.0}, cells={anchor={center}}, at={(0.02, 0.98)}, anchor={north west}}, axis background/.style={fill={rgb,1:red,1.0;green,1.0;blue,1.0}, opacity={1.0}}, anchor={north west}, xshift={1.0mm}, yshift={-1.0mm}, width={112.3mm}, height={61.5mm}, scaled x ticks={false}, xlabel={$N$}, x tick style={color={rgb,1:red,0.0;green,0.0;blue,0.0}, opacity={1.0}}, x tick label style={color={rgb,1:red,0.0;green,0.0;blue,0.0}, opacity={1.0}, rotate={0}}, xlabel style={at={(ticklabel cs:0.5)}, anchor=near ticklabel, at={{(ticklabel cs:0.5)}}, anchor={near ticklabel}, font={{\fontsize{10 pt}{13.0 pt}\selectfont}}, color={rgb,1:red,0.0;green,0.0;blue,0.0}, draw opacity={1.0}, rotate={0.0}}, xmode={log}, log basis x={10}, xmajorgrids={true}, xmin={10.0}, xmax={100000.0}, xticklabels={{$10^{1}$,$10^{2}$,$10^{3}$,$10^{4}$,$10^{5}$}}, xtick={{10.0,100.0,1000.0,10000.0,100000.0}}, xtick align={inside}, xticklabel style={font={{\fontsize{9 pt}{11.700000000000001 pt}\selectfont}}, color={rgb,1:red,0.0;green,0.0;blue,0.0}, draw opacity={1.0}, rotate={0.0}}, x grid style={color={rgb,1:red,0.0;green,0.0;blue,0.0}, draw opacity={0.1}, line width={0.5}, solid}, axis x line*={left}, x axis line style={color={rgb,1:red,0.0;green,0.0;blue,0.0}, draw opacity={1.0}, line width={1}, solid}, scaled y ticks={false}, ylabel={Impedance [$\Omega$]}, y tick style={color={rgb,1:red,0.0;green,0.0;blue,0.0}, opacity={1.0}}, y tick label style={color={rgb,1:red,0.0;green,0.0;blue,0.0}, opacity={1.0}, rotate={0}}, ylabel style={at={(ticklabel cs:0.5)}, anchor=near ticklabel, at={{(ticklabel cs:0.5)}}, anchor={near ticklabel}, font={{\fontsize{10 pt}{13.0 pt}\selectfont}}, color={rgb,1:red,0.0;green,0.0;blue,0.0}, draw opacity={1.0}, rotate={0.0}}, ymode={log}, log basis y={10}, ymajorgrids={true}, ymin={2.0e-10}, ymax={6.0e12}, yticklabels={{$10^{-9}$,$10^{-6}$,$10^{-3}$,$10^{0}$,$10^{3}$,$10^{6}$,$10^{9}$,$10^{12}$}}, ytick={{1.0e-9,1.0e-6,0.001,1.0,1000.0,1.0e6,1.0e9,1.0e12}}, ytick align={inside}, yticklabel style={font={{\fontsize{9 pt}{11.700000000000001 pt}\selectfont}}, color={rgb,1:red,0.0;green,0.0;blue,0.0}, draw opacity={1.0}, rotate={0.0}}, y grid style={color={rgb,1:red,0.0;green,0.0;blue,0.0}, draw opacity={0.1}, line width={0.5}, solid}, axis y line*={left}, y axis line style={color={rgb,1:red,0.0;green,0.0;blue,0.0}, draw opacity={1.0}, line width={1}, solid}, colorbar={false}]
    \addplot[color={rgb,1:red,0.0;green,0.6056;blue,0.9787}, name path={13}, draw opacity={1.0}, line width={1}, solid, mark={square*}, mark size={1.875 pt}, mark repeat={1}, mark options={color={rgb,1:red,0.0;green,0.0;blue,0.0}, draw opacity={1.0}, fill={rgb,1:red,0.0;green,0.6056;blue,0.9787}, fill opacity={1.0}, line width={0.75}, rotate={0}, solid}]
        table[row sep={\\}]
        {
            \\
            10.0  602.7623157810881  \\
            19.0  830.8505147319858  \\
            37.0  1159.4364658297227  \\
            72.0  1617.3810500808  \\
            139.0  2247.2615987710124  \\
            268.0  3120.437676434404  \\
            518.0  4338.543755729065  \\
            1000.0  6037.267916498533  \\
            1931.0  8788.32119109108  \\
            3728.0  26841.337076437172  \\
            7197.0  235502.54618245352  \\
            13895.0  2.330970242715848e6  \\
            26827.0  2.3260050223618414e7  \\
            51795.0  2.3247987695668283e8  \\
            100000.0  2.324465106356602e9  \\
        }
        ;
    \addlegendentry {$\|Z_{\mathrm{L}}\mathbf{I}_N+\mathbf{Z}_{\mathrm{R}}\|_{\mathrm{F}}$}
    \addplot[color={rgb,1:red,0.8889;green,0.4356;blue,0.2781}, name path={14}, draw opacity={1.0}, line width={1}, solid, mark={*}, mark size={1.875 pt}, mark repeat={1}, mark options={color={rgb,1:red,0.0;green,0.0;blue,0.0}, draw opacity={1.0}, fill={rgb,1:red,0.8889;green,0.4356;blue,0.2781}, fill opacity={1.0}, line width={0.75}, rotate={0}, solid}]
        table[row sep={\\}]
        {
            \\
            10.0  3.2924141261194785e-8  \\
            19.0  6.268325681679548e-8  \\
            37.0  1.221917606963265e-7  \\
            72.0  2.3789803761394968e-7  \\
            139.0  4.593906511060575e-7  \\
            268.0  8.858454515147464e-7  \\
            518.0  1.712307662983675e-6  \\
            1000.0  3.3057265083398364e-6  \\
            1931.0  6.383471053671013e-6  \\
            3728.0  1.2324079940778187e-5  \\
            7197.0  2.3792066671673894e-5  \\
            13895.0  4.5934636590847044e-5  \\
            26827.0  8.868586294510514e-5  \\
            51795.0  0.00017122627583864014  \\
            100000.0  0.0003305846787327021  \\
        }
        ;
    \addlegendentry {$L=15,\ \mathbf{u}=(10,1)\,\mathrm{m}$}
    \addplot[color={rgb,1:red,0.2422;green,0.6433;blue,0.3044}, name path={15}, draw opacity={1.0}, line width={1}, solid, mark={diamond*}, mark size={1.875 pt}, mark repeat={1}, mark options={color={rgb,1:red,0.0;green,0.0;blue,0.0}, draw opacity={1.0}, fill={rgb,1:red,0.2422;green,0.6433;blue,0.3044}, fill opacity={1.0}, line width={0.75}, rotate={0}, solid}]
        table[row sep={\\}]
        {
            \\
            10.0  1.6434379490572695e-9  \\
            19.0  3.123222941843172e-9  \\
            37.0  6.0827385693862515e-9  \\
            72.0  1.1837325991751944e-8  \\
            139.0  2.285323784922557e-8  \\
            268.0  4.4062972286029275e-8  \\
            518.0  8.516710554734534e-8  \\
            1000.0  1.644158728064218e-7  \\
            1931.0  3.1748766098758187e-7  \\
            3728.0  6.129441625421457e-7  \\
            7197.0  1.1833050993697148e-6  \\
            13895.0  2.284567006102943e-6  \\
            26827.0  4.41080155036401e-6  \\
            51795.0  8.515953429477596e-6  \\
            100000.0  1.644165217716491e-5  \\
        }
        ;
    \addlegendentry {$L=30,\ \mathbf{u}=(30,5)\,\mathrm{m}$}
\end{axis}
\end{tikzpicture}
            }
            \caption{Intra-array and inter-array coupling terms in~\cref{eq:inter_array_coupling} as functions of $N$ for fixed aperture $D=\qty{5}{\m}$, under two different propagation scenarios.}
            \label{fig:nlos_ula_aperture}
        \end{figure}
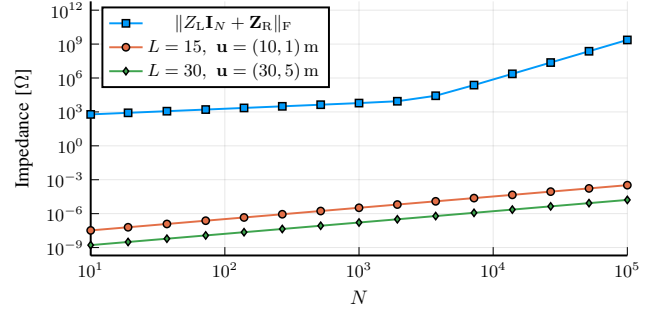

        The second scenario, corresponding to continuous arrays, is depicted in~\cref{fig:nlos_ula_aperture}.
        In this case, intra-array coupling initially also grows proportionally to $\sqrt{N}$; however, once $d \approx \lambda/20$, the growth rate increases substantially to $N^{3.5}$, as previously shown.
        Inter-array coupling, on the other hand, does not converge but grows linearly with $N$ over the entire range considered.

    \section{Discussion of Additional Scenarios}
        To conclude the technical developments presented in this paper, we discuss the applicability of the derived results to other scenarios, such as \acrshortpl{ula} with collinear dipoles and the downlink channel.

        \subsection{Collinear Dipoles}
            The mutual impedance between two collinear dipoles is given by
            \begin{equation}
                Z_{|}(r) = 3\R{r}\,\euler^{-\im \kappa r}\left(-\frac{1}{(\kappa r)^2} + \frac{\im}{(\kappa r)^3}\right),
            \end{equation}
            which decreases faster than \cref{eq:coupling_side_by_side} due to the toroidal radiation pattern of the dipoles~\cite[Sec.~4.2]{balanis_antenna_2016}.
            If we take into account that the coupling function of side-by-side dipoles converged to the generalized harmonic numbers in \cref{eq:bound_intra}, we can now conclude that collinear dipoles will exhibit the same divergence rate as side-by-side dipoles, driven by the $\sqrt{N}$ factor.

            Furthermore, since each dipole is electrically small, the same inter-array coupling model employed throughout the paper can also be assumed in this case.
            Therefore, the corresponding asymptotic analysis remains unchanged.

            Consequently, all results derived in this paper also apply to \acrshortpl{ula} composed of collinear Hertzian dipoles.

        \subsection{Downlink Channel}
            Throughout this paper, we have focused exclusively on the uplink channel, mainly because the results can be straightforwardly extended to the downlink.

            The physical channel, $\bsf{d}$, is reciprocal (up to a scaling factor), so the derived results apply directly without further modifications.
            Extending them to the information-theoretic channel, $\bsf{h}$, is also immediate, since in the absence of matching networks the uplink and downlink channels are reciprocal up to a linear transformation~\cite{damico_holographic_2024}.

            Therefore, the analysis carried out in this paper is equally applicable to the downlink channel.

        \subsection{Practical Antenna Arrays}
            As discussed in~\cref{sec:coupling_dipoles}, Hertzian dipoles are not physically realizable.
            Nevertheless, their mutual coupling exhibits qualitative behavior similar to that of half-wavelength dipoles~\cite[Sec.~8.6.2]{balanis_antenna_2016}, \cite{yordanov_arrays_2009,marti_physically_2026}.
            Moreover, inter-array coupling is evaluated in the radiative near field, where the point-source approximation applies.
            Consequently, the results presented in this work are expected to extend to practical antenna arrays beyond the Hertzian dipole model.

    \section{Conclusions}
        This paper shows that the unilateral approximation holds on average for massive \acrshort{simo} and \acrshort{miso} systems operating in near-field multipath environments modeled via correlated Rayleigh fading.
        Through asymptotic analysis, we demonstrate that for discrete arrays, intra-array coupling grows at a rate of $O(\sqrt{N})$ while inter-array coupling converges to a finite value, ensuring the approximation is met as $N\to\infty$.
        In the case of continuous arrays, the condition is also satisfied because intra-array coupling grows at $O(N^{3.5})$, significantly outpacing the $O(N)$ linear growth of inter-array coupling.
        Interestingly, these divergence rates coincide with those previously obtained for near-field \acrshort{los} propagation in~\cite{marti_asymptotic_2024}.

        As future research directions, the analysis could be extended to other array configurations such as planar arrays, as well as to more sophisticated antenna types.
        Furthermore, assessing the impact of the unilateral approximation on spectral efficiency would provide further insight into communication system performance.

    \bibliography{references}
\end{document}